\documentstyle[12pt]{article}
\begin{document}
\makeatletter
\makeatother
\begin{titlepage}
\font\fortssbx=cmssbx10 scaled \magstep2
\hbox to \hsize{
\hskip.5in \raise.1in\hbox{\fortssbx University of Wisconsin-Madison}
\hfill$\vcenter{\hbox{\bf MAD/PH/829}
                \hbox{\bf U\'{S}L-TH-94-01}
                 \hbox{May 1994}}$ }

\vspace{5 mm}
\begin{center}
{\Large {\bf {Phase Transition in Extended
Electroweak Theory.}}}\\
\vspace{7mm}
{\bf R.Ma\'{n}ka\footnote{ internet:
manka@usctoux1.cto.us.edu.pl} }\\
\vspace{3mm}
{\sl Department of Astrophysics and Cosmology,}\\
{\sl University of Silesia, Uniwersytecka 4, 40-007 Katowice,
Poland}\\
and\\
\sl Department of Physics\\
University of Wisconsin, Madison, WI 53706, USA
\end{center}
\setcounter{equation}{0}
\vspace{8 mm}
\centerline{ABSTRACT}
\vspace{2 mm}
The phase transition in the Weinberg-Salam-Glashow (GSW) electroweak
theory extended by the majoron and dilaton field is considered.
The possibility of the
boson  condensation in the extreme conditions in the standard
electroweak theory is shown. The first order phase transition
induces by the radiative corrections (the Coleman-Weinberg
potential) in the presence of matter was considered.
Due to t-quark mass ($\sim 174\ GeV$) a relatively high Higgs mass
($\sim 313 \ GeV$) was obtained. Only a fraction of this mass is
connected to the Coleman-Weinberg potential ($m_{CW} \sim 15\
GeV$). The model produces the first order phase transition for low
temperature ($T_c \sim 10\ GeV$).
Formation of bubbles filled with
matter was considered near the phase transition point.
The realistic ball with $ M \sim 10^5
- 10^9 M_{\odot}$
and the radius $ R \sim 10^{12} - 10^{14} cm$ is obtained.
\vfill
\leftline{PACS number(s): 98.80.Cq, 12.15.Cc}
\vspace{2 mm}
\end{titlepage}

\section{Introduction.}
The standard model describes the reality of the elementary
particle
interactions well known particularly in the perturbation sector.
The standard model was built in analogy to the Landau-Ginsberg theory of
superconductivity.
It is natural to expect a phase transition \cite{dkl:1972}
in similarity to superconductivity.
 If we believe, however that
the electroweak theory is really the nonabelian one, we should also
expect the nonperturbative effects  in extreme conditions (sufficiently
high temperature, high  matter densities or gravitational force).
One of the interesting features of the standard model is the scale
invariance in the high symmetric phase. This is anomalous symmetry and
quantum effects break it producing nonvanishing cosmological constant.
The classical scale invariance
 joins the standard model to
gravity.
In this paper the electroweak theory will be extended by the
dilatonic field and the singlet majoron field \cite{art:1994}.
The dilaton field appears in a natural way in
the Kaluza-Klein theories \cite{tap:1987}, superstring inspired
theories \cite{ew:1985}, \cite{sf:1989}  and in the
theories based on the noncommutative geometry approach \cite{ahc:1993}.
The spontaneous global lepton symmetry breaking leads to appearance of the
singlet majorana field \cite{ych:1981}
and the see-saw mechanism \cite{mgm:1991} of the neutrino mass
generation.\\
In this paper it will be shown that due to the dilaton field interaction both
standard model symmetry breaking scale and the global lepton symmetry breaking
scales are connected to each other. It will be shown that the electroweak
symmetry scale will be determined by the Coleman-Weinberg potential
coming with the quantum corrections to the standard electroweak
 theory.
\section{The theoretical background.}
The Glashow-Weinberg-Salam dilatonic model with $SU_{L}(2)\times
U_{Y}(1)$
symmetry
is described by the Lagrange function
\begin{equation}
{\cal L} = {\cal L}_b + {\cal L}_f
\end{equation}
\begin{eqnarray}
{\cal L}_b &=&  -\frac{1}{4} e^{-2 \kappa \varphi (x) } F^{a}_{\mu
\nu}F^{a\mu \nu} -
\frac{ 1}{4 } e^{-2 \kappa \varphi (x) } B_{\mu \nu}B^{\mu \nu}
\\ \nonumber
& + &
\frac{ 1}{2 }{\partial}_{\mu} \varphi (x) {\partial}^{\mu}
\varphi (x) +
(D_{\mu}H)^{+}D^{\mu}H  +
\frac{ 1}{2 }{\partial}_{\mu} \chi (x) {\partial}^{\mu} \chi (x) - U(H)-U(\chi)
\end{eqnarray}
with the $SU_{L}(2)$ field strength tensor
\begin{equation}
F^{a}_{\mu \nu} = \partial_{\mu}W^{a}_{\nu} -
\partial_{\nu}W^{a}_{\mu} +g\epsilon_{abc}W^{b}_{\mu}W^{c\nu}
\end{equation}
and the $U_{Y}(1)$ field tensor
\begin{equation}
B_{\mu \nu} = \partial_{\mu}B_{\nu} - \partial_{\nu}B_{\mu}.
\end{equation}
The covariant derivative is given by
\begin{equation}
D_{\mu} = \partial _{\mu} -\frac{1}{2}igW^{a}_{\mu}\sigma^{a}-
\frac{1}{2}g^{'}YB_{\mu}
\end{equation}
where $B_{\mu}$ and
\begin{equation}
W_{\mu}=\frac{ 1}{2 } W^{a}_{\mu}\sigma^{ a}
\end{equation}
are a local gauge fields
associated with the
$U_{Y}(1)$ and $SU_L(2)$ symmetry group, respectively. $Y$ is a
hypercharge. The gauge group is simply the
multiplication of $U_{Y}(1)$ and $SU_{L}(2)$ so there are two
gauge
couplings $g$ and $g'$.
Generators of the gauge groups are unit matrix for $U_{Y}(1)$ and
Pauli matrixes for $SU_{L}(2)$.
In the simplest version of the standard model a doublet of Higgs
field is introduced
\begin{equation}
 H = \left( \begin{array}{c}
H^{+} \\ H^{0}
\end{array}   \right)
\end{equation}
with the Higgs potential
\begin{equation}
U(H) = \lambda(H^{+}H - \frac{1}{2}v^{2}_0 e^{-2
\kappa
\varphi (x) })^{2}
\end{equation}
The form of the potential leads to a degeneracy of the vacuum and
to a
nonvanishing vacuum expectation value of the
Higgs field
and in consequence to the fermion and boson masses.
The similar type of potential we may expect for the complex field
$\chi$
\begin{equation}
U(\chi) = \lambda_S({\chi}^{+}\chi - \frac{1}{2}u^{2}_0 e^{-2
\kappa
\varphi (x) })^{2}
\end{equation}
$U(\chi) $ has the global lepton U(1) symmetry.\\
The fermion contents of the model is extended only by the right handed
neutrino $ \nu_R$ as a singlet of the $SU_L(2)\times U(1)$ group.
For simplicity, let us limit ourselves to the first lepton family.
We have the lepton lagrangian as follows:
\begin{eqnarray}
{\cal L}_f = ie^{+}_R \sigma^{\mu}\partial_{\mu}e_R +i\nu^{+}_R
\sigma^{\mu}\partial_{\mu} \nu_R
+ i L^{+}\sigma^{\mu}D_{\mu} L +\\ \nonumber
 i h_e (HLe_R +h.c)+
ih_{\nu}(L^{+}\epsilon H\nu_R + h.c.)+
ih_R ( {\nu}^2_R \chi + h.c),
\end{eqnarray}
where $\sigma^{\mu}=\{ I,\sigma^i \}$
Here we also adopted the notation
\begin{equation}
L = \left( \begin{array}{c}
\nu_{L} \\ e_{L}
\end{array}   \right),L=e,\nu,\tau
\end{equation}
\\
Let us consider a system of quantum boson fields
\begin{equation}
\phi_{A} = \{ \varphi, W^{a}_{\mu} , B_{\mu}, H, \chi \}
\end{equation}
\begin{equation}
{\Phi}_{ A}= \sum_{ \lambda}\frac{ 1}{\sqrt{ (2 \pi)^{ 3}} } \int
\frac{ d^3 k}{\sqrt{ 2 \omega_{ k}} } \{a_{ {\bf k}, \lambda}e^{
i{\bf k x} -i \omega_{ k}t }+ a^{+ }_{ {\bf k}, \lambda}e^{ -
i{\bf k x} + i \omega_{ k}t } \}
\end{equation}
\begin{equation}
[a_{ {\bf k}, \lambda},a^{+ }_{ {\bf k}', \lambda'}]=
g_{\lambda,\lambda' } \delta ({\bf k} -{\bf k}')
\end{equation}
with the vacuum state $|0>$ defined as $ a_{ {\bf k}, \lambda}|0>
=0$
 and a new
system of quantum boson fields $\tilde{\phi}_{A}$ related to
$\phi_{A}$
by
\begin{equation}
\label{row_84}
\phi_{A} = \tilde{\phi}_{A} + \xi_{A} \; ,
\end{equation}
where the shifts $\xi_{A}$ are the classical fields.
These shift transformations can be expressed as:
\begin{displaymath}
\label{row_85}
\tilde{\phi}_{A} = {\cal D}(\xi_{A}) \phi_{A} {\cal
D}^{+}(\xi_{A}) \; ,
\end{displaymath}
where
\begin{displaymath}
\label{row_86}
{\cal D}(\xi_{A}) = exp \sum_{A} \sum_{\lambda} \int d^{3}k
(\xi_{A {\bf k}\lambda} a_{A {\bf k} \lambda}^{+} - \xi^{*}_{A {\bf
k}\lambda }
a_{A {\bf k} \lambda}) \; .
\end{displaymath}
Here $\sum_{A}$ is the sum over all shifted fields and
$\sum_{\lambda}$ means
the sum over all degrees of freedom for these fields. The $a_{A
{\bf k} \lambda}$
and  $a_{A {\bf k}  \lambda}^{+}$  are  the   annihilation   and
creation
operators for the $ \phi_{A}$ field.
The coefficients $\xi_{A {\bf k}\lambda}$ are the Fourier
transformations of
the $\xi_{A}$ fields.
Now we assume that in the Hilbert space ${\cal H}$ there exists
a normalized
vacuum vector $\mid 0>$ which is annihilated by the operators
$a_{A {\bf k} \lambda}$
\begin{displaymath}
\label{row_87}
a_{A {\bf k} \eta} \mid 0> = 0 \; and \; <0 \mid 0> = 1 \; .
\end{displaymath}
The shifts cause the changing of the ground state of a system
according to
the relation:
\begin{displaymath}
\label{row_88}
\mid 0> \longrightarrow \mid \tilde{0}> = {\cal D}(\xi_{A}) \mid
0> \; .
\end{displaymath}
The new vacuum state $\mid \tilde{0}>$ is simply the Glauber
coherent state.
This state includes the infinite number of excited states of
$\phi_{A}$
fields. The state $\mid \tilde{0}>$ is also normalized, i.e.,
$<\tilde{0}
\mid \tilde{0}> = 1$.
As the state $\mid 0>$ is the vacuum state for the $\phi_{A}$
fields also the
state $\mid \tilde{0}>$ may be considered as the vacuum state for
the
$\tilde{\phi}_{A}$ fields.
Hence, when we have $<0 \mid \phi_{A} \mid 0> = 0$ we also have
$<\tilde{0}
\mid \tilde{\phi}_{A} \mid \tilde{0}> =0$ and
\begin{displaymath}
\label{row_89}
<\tilde{0} \mid \phi_{A} \mid \tilde{0}> = <\tilde{0} \mid
\tilde{\phi}_{A}
\mid \tilde{0}> + \xi_{A} = \xi_{A} \; .
\end{displaymath}
The point is that when the ground state $\mid \tilde{0}>$ is
attained as
the result of the transformation  which is
not the gauge
symmetry transformation or as the result of the appearance of
some new
external charges in the system it leads to the conclusion that
the Fock
spaces which are built on the ground states $\mid 0>$ or $\mid
\tilde{0}>$,
respectively, are not unitary equivalent. This means that some
classical
boson fields $\xi_{A}$ may attain physical interpretation.
The physical system is totally defined by the free energy
\begin{equation}
F=-kT ln Tr(e^{ -\beta H})
\end{equation}
where  H is hamiltonian of the physical system
\begin{equation}
H=\sum_A \int d^3 x \{ \partial_0 \Phi_A \pi^A_{ \Phi}  -  {\cal L} \}
\end{equation}
and  $ \pi^A =\frac{ \partial{\cal
L}}{\partial(\partial_{ 0}\Phi_A)
}$
is a momentum connected to $\Phi_A$.
In this paper we shall use the effective potential approach built
using the Bogolubov inequality \cite{rm:1986}
\begin{equation}
F \leq  F_1 = F_{ 0 }(m^2)+ < H - H_{ 0}>_{ 0}
\end{equation}
$F_0$ is the free energy of the trial system
\begin{equation}
F_{ 0}=U_{CW} + \sum_A \{ \frac{ 1}{24 }m^2_A T^2 - \frac{ 1}{12 \pi } m^3_A T
- \frac{
m^4_A}{64\pi^2 }ln(\frac{m^2_A }{cT^2 }) \} + ...
\end{equation}
with $ c=\frac{ 3}{2 }+2 ln (4 \pi) -2 \gamma \cong 5.4 \ $.
$U_{CW}$ is the Coleman-Weinberg potential.
The hamiltonian of the system is defined as usual as
\begin{equation}
H=\sum_A \int d^{ 3}x (\pi_{\Phi_A }\dot{\Phi^A}-{\cal L})
\end{equation}
The trial system we shall suppose as effectively free quasiparticle
system described by the Lagrange function
\begin{equation}
{\cal L}_{ 0}=\sum_A \frac{ 1}{2 }\partial_{
\mu}\tilde{\Phi_A}\partial^{ \mu}\tilde{\Phi_A}-\frac{
1}{2 }m^2_A \tilde{\Phi_A}^2
\end{equation}
We decompose the $ \Phi_A$ field into two components, the
effectively free quasiparticle field $ \tilde{\Phi_A}$ and the
classical boson condensate $ \xi_A$
\begin{equation}
\Phi_A =\tilde{\Phi_A} + \xi_A
\end{equation}
The $\xi_A $ field will be treated as the variational parameters
in the effective potential.
\section{\bf The electroweak phase transition.}
The standard model was built in analogy to the Landau-Ginsberg theory of
superconductivity where is the continuous phase transition.
Indeed, in the first approximation we have only  condensation of the Higgs
field.
\begin{equation}
 H = \left( \begin{array}{c}
H^{+} \\ \frac{ 1}{\sqrt{ 2} } v
\end{array}   \right)= \tilde{H} + \left( \begin{array}{c}
0 \\ \frac{ 1}{\sqrt{ 2} } v_0
\end{array}   \right)
\end{equation}
This happens if we neglect the radiative corrections giving the
Coleman-Weinberg potential. Including only temperature effects the
minimal standard electroweak model has the effective potential
\begin{equation}
U_{ eff}=\frac{1}{2}DT^2 v^2+ \frac{ 1}{4 }  \lambda ( v^2 -v^2_0)^2
\end{equation}
with
\begin{equation}
D= \sum_A \frac{1}{12}(\frac{m_A}{v_0})^2 =
\frac{1}{12} \{ (\frac{m_W}{v_0})^2+ (\frac{m_Z}{v_0})^2+
(\frac{m_H}{v_0})^2 \}
\end{equation}
This potential has temperature dependent minimum
\begin{equation}
v^2_T = v^2_0 - (\frac{D}{\lambda})T^2
\end{equation}
which vanishes at phase transition point
\begin{equation}
T_c =\sqrt{\frac{D}{\lambda}} v_0 = \frac{1}{2 D}m_H
\end{equation}
This phase transition temperature is really high, for example for
$m_H = 302 \ GeV$ we have $T_c = 464.6 \ GeV$.\\
In the extended standard model we may expect condensation of the following
fields
\begin{equation}
\varphi =\tilde{\varphi} + \sigma \ \ \ \ \ \  or \ \ \ \ \
\ D= e^{ -\kappa \sigma}
\end{equation}
\begin{equation}
W^{ a}_{ \mu }=\tilde{ W}^{ a}_{ \mu}+ {\it a}^{ a}_{ \mu}
\end{equation}
\begin{equation}
B_{ \mu }=\tilde{ B}_{ \mu}+ {\it b}_{ \mu}
\end{equation}
\begin{equation}
 H = \left( \begin{array}{c}
H^{+} \\ \frac{ 1}{\sqrt{ 2} } v
\end{array}   \right)= \tilde{H} + \left( \begin{array}{c}
0 \\ \frac{ 1}{\sqrt{ 2} } v_0
\end{array}   \right)
\end{equation}
\begin{equation}
\chi =\tilde{\chi} + \frac{ 1}{\sqrt{ 2} } u_0=\frac{ 1}{\sqrt{ 2} }
( u + i \varphi_M )
\end{equation}
with $\tilde{\chi}= \frac{ 1}{\sqrt{ 2} } ( \varphi_u +i \varphi_M )$.
$ \varphi_M $ is the majoron field.
In the result of bosons condensation the Higgs mechanism generates
 not only the Dirac mass
\begin{equation}
m_{D\nu} = \frac{1}{\sqrt{2}}h_{e}v_0
\end{equation}
but the lepton number violating Majorana mass
\begin{equation}
M = \frac{1}{\sqrt{2}}h_{\nu}u_{0}
\end{equation}
as well.
Thus the neutrino mass matrix can be written as follows
$$
{\cal M} =
\pmatrix{0 & m_{D\nu} \cr m_{D\nu}  & M }
$$
In the case $M=0$ only the Dirac neutrino may be obtained.
In general, it should have the same mass as the electron or quark
($ \sim 1\  MeV$). In the broken phase  due to the see-saw
mechanism we obtain two Majorana mass eigenstates \cite{bb:1992}
 \begin{equation}
m_{\nu,M} =\frac{ 1}{2}M \{1 \mp \sqrt{1+4 ({\frac{
m_{D\nu}}{M}})^2} \}
\sim \{ -\frac{{m_{\nu,D}}^2}{M} , M\}
 \end{equation}
The astrophysical boundaries \cite{cko:1993} suggest that $u_0 \ \sim \ v_0$
and that Yukawa coupling $h_{\nu} = 10^{-17}$ is small. This gives the mass
of the Dirac neutrinos  $m_{\nu,D} \sim 25  KeV$
If we estimate $ M \sim 100 $ GeV we have
$ m_{1,\nu,M} \sim 2.7\ 10^{-3}$ eV and $ m_{2,\nu,M} \sim  100$ GeV for the
Majorana neutrinos.
The classical potential U gives
\begin{eqnarray}
U_{ 0}(v,\sigma) =
\frac{ 1}{4 }\lambda ( v^2 -v_{ 0}^2 e^{-2\kappa \varphi (x) } )^2 +
\frac{ 1}{4 }\lambda_S ( u^2 -u_{ 0}^2 e^{-2\kappa \varphi (x) } )^2\\
\nonumber
=\frac{ 1}{4 }\lambda ( v^2 -v_{ 0}^2 D^2 )^2 +
\frac{ 1}{4 }\lambda_S ( u^2 -u_{ 0}^2 D^2 )^2
\end{eqnarray}
\begin{equation}
\frac{ \partial U_{ 0}}{\partial v }=0 \ \ \ gives
\ \ \ \ \ \lambda ( v^2 - v^2_{ 0}D^2 ) v =0
\end{equation}
\begin{equation}
\frac{ \partial U_{ 0}}{\partial u }=0 \ \ \ gives
\ \ \ \ \ \lambda ( u^2 - u^2_{ 0}D^2 ) u =0
\end{equation}
\begin{equation}
\frac{ \partial U_{ 0}}{\partial D }=0 \ \ \ gives
 \ \ -\lambda ( v^2 - v^2_{ 0} D^2 ) v^2_{ 0}D -
\lambda_S ( u^2 - u^2_{ 0} D^2 ) u^2_{ 0}D=0
\end{equation}
Apart of the trivial solution  $v=0,\ D=0 $ we have
\begin{equation}
D^2 =\frac{ v^2}{v^2_{ 0} }
\end{equation}
and
\begin{equation}
u^2 =u^2_0 D^2 = \frac{ u^2_0}{v^2_{ 0} }v^2
\end{equation}
It is interesting that in the presence of the dilaton
$\varphi(x)$
the classical potential vanishes at the minimum point
\begin{equation}
U_{ 0}(v)=U_{ 0} (v,u=(u_0/v_0)v,D=v/v_{ 0})=0
\end{equation}
There is no cosmological term on the classical level.
Now we can define the standard model Higgs field $\varphi_v $,
Higgs field $\varphi_u $ connected to the global $ U_L (1)$ symmetry
 and dilaton
field $\varphi_d$ as
\begin{equation}
v=v_0 + \varphi_v
\end{equation}
\begin{equation}
u=u_0 + \varphi_u
\end{equation}
\begin{equation}
d=v_0 + \kappa v_0 \varphi_d
\end{equation}
The Higgs field mass is determined from the ``mass matrix"
\begin{equation}
m^2_{ i,j}|_{min}  =
 \frac{ \partial^2 U_{ 0}}{\partial \Phi_{
i}\partial\Phi_{ j}} \\ \nonumber
\end{equation}
\begin{eqnarray}
m^2_{ i,j}|_{min}  =
 \left( \begin{array}{c}
2 \lambda v_0^2 ,\ \ \ \ \ \ 0,
\ \ \ \ \ \ \ \ -2 \lambda v^2_0 (\kappa v_0)\\
0, \ \ \ \ \ \ 2 \lambda_S u_0^2 ,
\ \ \ \ \ \ -2 \lambda_s u^2_0 (\kappa v_0)\\
-2 \lambda v^2_0 (\kappa v_0), -2 \lambda_s u^2_0 (\kappa v_0),
 2 (\kappa v_0)^2 (\lambda v_0^2  +\lambda_S (\frac{u_0}{v_0})^2 u_0^2  )
\end{array}   \right)
\end{eqnarray}
where now the classical fields are $ \Phi=( \varphi_v,d=\varphi_v)$.
At the extremum point  the diagonalized mass matrix has the form
\begin{equation}
diag \ \ m^2_{ i,j}|_{min}= \ \{ m^2_{ H}=2\lambda v^2,
m^2_{L}= \lambda_S u^2_0,
 m_{ d}=0 \}
\end{equation}
The physical fields are result of diagonalization of this mass matrix.
They may be defined as $\Phi_{i,ph}=(\varphi_H,\varphi_L, \varphi_D )$
The physical fields are an orthogonal mixture
$\Phi_{i,ph}=(\varphi_u, \varphi_u,\varphi_d )$
\begin{equation}
\Phi_{i,ph}= R^j_i \Phi_j
\end{equation}
where $R^j_i$ is orthogonal matrix diagonalizing the mass matrix (47).
As $(\kappa v_0) \ \sim \ 10^{-17}$ is really very small number, this
mixing is very small. $\varphi_H \sim \varphi_v$ is a
standard model Higgs particle, $\varphi_L \sim \varphi_u$ is a Higgs particle
connected to the spontaneous $U_L(1)$ symmetry breaking, $\varphi_D$ is the
dilaton field.
We have  not determined on the classical level. It must be
determined by the radiative corrections --- the Coleman-Weinberg
effective potential \cite{msh:1989}.
The gauge field condensation
\begin{equation}
{\it a}^{a}_{ \mu}= \{ {\it a}^{a}_{ 0}= \zeta {\delta}^{a}_{ 0},
{\it
a}^{a}_{ i}=0 \}
\end{equation}
\begin{equation}
{\it b}_{ \mu}= \{ {\it b}_{ 0}= \eta , {\it b}_{ i}=0 \}
\end{equation}
For example we have
\begin{equation}
D_{ \mu}H^{ +}D^{ \mu}H = \frac{ 1}{4 }g^2 v^2 \zeta^2 + \frac{
1}{4
}g'^2 v^2 \eta^2 + \frac{ 1}{2 }M^2_{ Z} Z_{ \mu}Z^{ \mu} +
\frac{ 1}{8 }g^2 v^2 \frac{ g'}{\sqrt{ g^2 +g'^2} } {\it a}^3_{
0
}Z^{ 0} \ \ + ...
\end{equation}
We have used redefinition of the gauge field
$$
\left( \begin{array}{c}
W^{ 3}_{ \mu} \\ B_{ \mu}
\end{array}   \right) =\left( \begin{array}{c}
cos \vartheta_{ W}    \ \   sin \vartheta_{ W}  \\ -sin \vartheta_{W}
\ \  cos \vartheta_{ W} \end{array}   \right) \left( \begin{array}{c}
Z^{ 3}_{ \mu} \\ A_{ \mu} \end{array}   \right)
$$
with
$$
cos \vartheta_{ W}= \frac{ g}{\sqrt{ g^2 + g'^2} }
$$
the $W$ and $Z$ bosons masses are the same as in the standard model
$$
M^2_{ W} = \frac{ 1}{4 } g^2 v^2
$$
$$
M^2_{ Z} = \frac{ 1}{4 } (g^2 +g'^2) v^2
$$
$$
Z_{ \mu}={\tilde{Z}}_{ \mu} + {\it z}_{ \mu}
$$
$$
A_{ \mu}={\tilde{A}}_{ \mu} + {\it a}_{ \mu}
$$
where
$$
{\it a}_{ 0}= cos \vartheta_{ W} \zeta - sin \vartheta_{ W} \eta
$$
If we do not want to break the $U_Q (1)$ electromagnetic gauge
symmetry, we should impose the condition
\begin{equation}
{\it z}_0 = \frac{ \eta}{cos \vartheta_{ W} }
\end{equation}
This gives
\begin{equation}
{\cal L} = \frac{ 1}{2 } M^2_{Z}(v) {\it z}^2_0 + \rho_Z {\it
z}_{
0} \ + \ ...
\end{equation}
where
\begin{equation}
 \rho_Z \Leftarrow {\cal J}^{ \mu}_{ Z} ={\cal J}^{3, \mu}_{ W}
cos
\vartheta_{ W} + {\cal J}^{ \mu}_{ Y} sin \vartheta_{ W} +
\end{equation}
\begin{equation}
\frac{ \partial{\cal L}}{\partial {\it z}_0 }=0 \ \ \ \ gives \
\ \ \ \ {\it z}_0 = - \frac{ \rho_Z}{M^2_Z (v)}
\end{equation}
In the presence of the weak external neutral charge $ \rho_Z$ we
have an additional term
\begin{equation}
U_{add} = \frac{ 1}{2 }  \frac{ \rho^2_Z}{M^2_Z (v)}
\end{equation}
\begin{equation}
U(v) = U_{ add}(v)+ U_{ CW}(v)
\end{equation}
where the quantum corrections ($ \sim \hbar$) generate the Coleman
- Weinberg potential
Keeping only the contributions associated with the gauge bosons
$W$,
$Z$ and the
quark $t$  the radiative corrections give
 \begin{equation}
 U_{ CW}(v) = \sum_{ i=\varphi_H, W,Z,t,...}\frac{ n_{ i}}{64 \pi^2 } m^{
4}_{
i}(v)
 \{ ln \frac{ m^{ 4}_{ i}(v) }{Q^2 }-\frac{ 3}{2 } \}
  \end{equation}
  where for example
 \begin{equation}
 m^2_W =\frac{ 1}{4 }g^2 v^2 ,
  \end{equation}
  \begin{equation}
m^2_Z =\frac{ 1}{4 }(g^2+g'^2) v^2 ,
 \end{equation}
  \begin{equation}
  m^2_t = h^2_t v^2, ...
   \end{equation}
$ n_i$ depends on the number of degrees of freedom and the
particle's
statistics
 \begin{equation}
n_{\varphi_H}=1, \  n_W=6, \ n_Z=3, \ n_t =-12, ...
  \end{equation}
$Q$ is the renormalization scale. Let us notice that $ V_r
(0)=0$.
In our calculation we should also include the contribution from
the $U_L(1)$ Higgs field and from the right handed neutrino $\nu_R$.
They have masses
\begin{equation}
m^2_L \sim 2\lambda_S (\frac{u_0}{v_0})^2 v^2
\end{equation}
\begin{equation}
m^2_{\nu_R}  = h^2_{\nu_R} (\frac{u_0}{v_0})^2 v^2
\end{equation}
Their number of degrees of freedom are
\begin{equation}
n_L =1, n_{\nu_R}=2*3
\end{equation}
where 2 is the spin degree of freedom and 3 is the number of fermion
families.
The Coleman-Weinberg potential may be written in the form
\begin{equation}
U_{ CW}(x)= \frac{ 1}{4 }C v^4 \{ ln\frac{ v^2}{Q^2 }-\frac{
25}{6 } \}
\end{equation}
where
\begin{equation}
C=\frac{ 1}{16 \pi^2 }\{ 6 (\frac{m_W }{v_0 })^4 + 3(\frac{m_Z
}{v_0 })^4 + (\frac{m_H}{v_0})^4
+ (\frac{m_L }{v_0 })^4 -12 (\frac{m_t }{v_0 })^4 -6 (\frac{m_{\nu_R} }{v_0
})^4 \}
\end{equation}
The minimum point $v_0$ define
\begin{equation}
v_0 = Q e^{\frac{ 11}{6 }}
\end{equation}
Finally we have
\begin{equation}
U_{ CW}= \frac{ 1}{4 } C v^4 \{ ln (\frac{ v^2}{v^2_{ 0} }) -
\frac{
1}{2 } \}
\end{equation}
It is interesting that the radiative corrections produce a negative
cosmological constant
\begin{equation}
B = - \frac{ 1}{128 } \{ 6 m^4_W + 3 m^4_Z + m^4_H+m^4_L- 12 m^4_t
-2m^4_{\nu_R}\}
\end{equation}
This puts the cosmological boundaries on the quark top mass.
For example for $ M_t \sim \ 174 \ GeV$ \cite{chi:1994}
$m_{\nu_R}=120 \ GeV$ and $m_L=250\ GeV$ we have
$ m_H \ > \ 104 \ GeV $ and  B = $ 10^{ 5} \ \ GeV^4$.
When we include the Coleman-Weinberg potential the mass matrix will
change a bit
\begin{equation}
m^2_{ i,j}|_{min}  =
 \frac{ \partial^2 U_{ 0}}{\partial \Phi_{
i}\partial\Phi_{ j}} \\ \nonumber
\end{equation}
\begin{eqnarray}
m^2_{ i,j}|_{min}  =
 \left( \begin{array}{c}
2 \lambda v_0^2+\frac{{\partial}^2 U_{CW}}{\partial v^2} ,\  0,
\ \ \ \ \ \ \ \ -2 \lambda v^2_0 (\kappa v_0)\\
0, \ \ \ \ \ \ 2 \lambda_S u_0^2 ,
\ \ \ \ \ \ -2 \lambda_s u^2_0 (\kappa v_0)\\
-2 \lambda v^2_0 (\kappa v_0), -2 \lambda_s u^2_0 (\kappa v_0),
 2 (\kappa v_0)^2 (\lambda v_0^2  +\lambda_S (\frac{u_0}{v_0})^2 u_0^2  )
\end{array}   \right)
\end{eqnarray}
For small $(\kappa v_0) \sim 10^{-17} $ the Higgs particle mass is equal to
\begin{equation}
m^2_H = m^2_0 + m^2_{CW}
\end{equation}
with tree level Higgs mass
 \begin{equation}
m^2_0 = 2 \lambda v^2_0
\end{equation}
and the Coleman-Weiberg  Higgs mass
 \begin{equation}
m^2_{CW}= \frac{{\partial}^2 U_{CW}}{\partial v^2} = 2 C v^2_0
\end{equation}
The dilaton mass is equal to
\begin{equation}
m^2_D = \frac{1}{2}m^2_0 (\kappa v_0)^2 ( 1- \frac{m^2_{CW}}{m^2_0+m^2_{CW}})
\ \ \ \sim \ \ \ 10^{-6}\ eV
\end{equation}
Temperature contributions   to the effective potential
originated from $F_0$
 may also be included.
At last the effective potential has the form
\begin{equation}
U_{ eff}=\frac{1}{2}DT^2 v^2+ \frac{ 1}{4 } C v^4 \{ ln (\frac{ v^2}{v^2_{ 0}
}) -
\frac{
1}{2 } \} + U_{add}
\end{equation}
with
\begin{equation}
D= \sum_A \frac{1}{12}(\frac{m_A}{v_0})^2 =
\frac{1}{12} \{ (\frac{m_W}{v_0})^2+ ...
\end{equation}
Let us neglect for the moment $U_{add}$.  The extremum $U_{eff}$ points
$v_T$ obey the equation
\begin{equation}
DT^2 +  Cv^2_T ln(\frac{v^2_T}{v^2_0}) = 0
\end{equation}
At the first order phase transition point $T_c$ we have the degenerate
values of the free energy ( effective potential). This means that
\begin{equation}
U_{eff}(0) = U_{eff}(v_c)
\end{equation}
with $v_c=v_{T_c}$. This condition defines the first order phase transition
temperature $T_c$
\begin{equation}
T^2_c = \frac{C}{2 D} v^2_c \sim \frac{C}{2 D} v^2_0
\end{equation}
The  known particles masses allows us to establish boundaries on Higgs particle
mass and quark top mass \cite{wbu:1990} $(c>0)$ and the phase transition
temperature. For example in the minimal standard model (without dilaton and
majoron field) we have $m_t < 100\ GeV$. For two Higgs doublet like in the
supersymmetrical extension we have $m_t < 100\ GeV$. In the model built on
noncommutative geometry there is limitation   $    m_t \sim 130 GeV$.
In the extended model if $m_t > \ 89\ GeV$ then Majorana neutrino mass
$m_i <\  130 GeV$.  All these estimations allows us to predict the
 phase transition temperature  [Table 1]
\begin{equation}
T_c \ \sim \ 10\mbox{--}30\ GeV
\end{equation}
It is rather low temperature in comparison to the minimal standard model
($ T_C = 464.6 GeV$ for $m_H = 302 GeV$).

\section{\bf The astrophysical meaning.}
Let us consider now the  nonhomogeneous Higgs field  configuration
\cite{dr:1992}
near the first order phase transition point. In the presence of
fermions  we shall have two ($v_{ *},v $) different from zero
minima (Fig.~1). These two minima are degenerate at the phase
transition point $T_{ c} \sim 10 -20 GeV$. When $T_{ c} \rightarrow
0 $ we have the limit ($ v_{ *}=0, v=v_0$). Let us now define the
effective  field
\begin{equation}
 \Phi = \frac{ 1}{\sqrt{ 2} }(v - v_{ *} )=\frac{ 1}{\sqrt{ 2} }x
\end{equation}
Effectively, the Higgs field may be described near the phase
transition point as
 \begin{equation}
{\cal L}=\partial_{\mu}{\Phi}^{+} \partial^{\mu}\Phi -U(\Phi)
 \end{equation}
with
 \begin{equation}
U(\Phi)= \lambda_{*
}(\Phi^{+}\Phi)|\Phi-\frac{1}{\sqrt{2}}x_0|^{2}
 \end{equation}
The parameter $\lambda_{*}$ determines the potential wall height
between two minima. In the first approximation
\begin{equation}
U_0 = \frac{ 1}{16 } \lambda_{*} v^4_0
\end{equation}
The Lagrange equation gives
\begin{equation}
\Box x=\frac{\partial U}{\partial x}
\end{equation}
In the spherical coordinates this equation takes the form
\begin{equation}
\frac{d^2 x}{dr^2} +\frac{2}{r} \frac{dx}{dr}=
\frac{\partial U}{\partial x}
\label{wz_2}
\end{equation}
As the potential takes
the degenerate form
\begin{equation}
U=\lambda{'} x^2 ( x -x_0)^2
\end{equation}
where $x_{0}$ is obtained from the Coleman-Weinberg potential
$U_{CW}$.
In the thin wall approximation we neglect the second term in
equation (\ref{wz_2}).
As a result we obtain a one dimensional equation which is easy to
solve.
\begin{equation}
\frac{1}{2} (\frac{dx}{dr})^2 = U
\end{equation}
In this approximation the solution may be described as the ball
with the radius $R$
  \begin{equation}
x       =        \left \{ \begin{array}{ll}
\ \ \ \ \  \ \ \ \ \ \ \ \ 0\ \  \ \ \ \ \ \ \ \ \ \ \ \ \ \ \ \ \
\  \ r \le R \\
 x_0 e^{m_{\Phi}(r-R)} /(1+e^{m_{\Phi}(r-R)}) \ \ r>R
                \end{array}
                \right \}
                \label{wz_3}
         \end{equation}
where
\begin{equation}
m^2_{\Phi} =\frac{ \partial U_{ef}}{dx } |_{x_0}
\end{equation}
defines the scalar Higgs field mass. Its inverse $ l =1/m$ is the
coherent
length and measures the ball wall size. So, the wall is really thin
and this
approximation
seems to be reasonable. The scalar ball may be thought of as a
constant
solution
inside the ball and the soliton solution representing the wall.
{}From the solution (\ref{wz_3}) one can conclude that inside the
ball for $r < R$
$x = 0$. That means that inside the ball exists a
 phase  with the gauge boson ($Z_{ \mu}$) condensation.
Outside the soliton $(x \neq 0)$ we have the low symmetry phase
with the broken electroweak symmetry.
In this region all fermions get masses.
Because inside the ball all fermions are nearly massless whereas
outside  they get
large masses, they have the natural tendency to fill the ball.
They will give the stabilizing (repulsive) term in the expression
for the
total energy of the whole system which will protect from the
gravitational
collapse.
The boson part of the ball energy
 \begin{equation}
{\cal E}_b =\frac{ 4\pi}{3 }B+4\pi \int^{\infty}_R drr^2
\{ \frac{ 1}{2 }(\frac{ dx}{dr })^2 +U_{ef}(x) \}
 \end{equation}
In the thin wall approximation we have
  \begin{equation}
           {\cal E}_b= s R^2+\frac{ 4\pi}{3 }B R^3
\end{equation}
where
\begin{equation}
s \sim \frac{ \sqrt{ 2 \lambda_{ *}}}{8 } x^3_0
\end{equation}
is the surface tension.
The fermion energy corresponds to the repulsive force coming from
the
Pauli principle
\begin{equation}
{\cal E}_f =\frac{AN^{\frac{ 4}{3}}}{R}
\end{equation}
with
\begin{equation}
A =\frac{4}{3}{\pi}^{2}(\frac{9\pi}{2})^{\frac{4}{3}}
{\gamma}^{-\frac{1}{3}}
\end{equation}
where $\gamma$ is a number of degrees of freedom.
The total energy of the ball is equal
\begin{equation}
{\cal E}=\frac{N^{\frac{ 4}{3 }}A}{R} +s R^2+\frac{ 4\pi}{3 }B R^3
\end{equation}
In this model we consider the case when the bag constant $B = 0$ so
in the
expression for total energy remains only the term containing the
surface
tension $s$.
\begin{equation}
{\cal E}=\frac{N^{\frac{ 4}{3 }}A}{R} +s R^2
\end{equation}
For example, for $ x_0 \sim v_0= 246 GeV
s= 2.7 \ 10^5 (GeV)^3$.
Minimizing, $\frac{\partial {\cal E}}{\partial R} =0$ gives
\begin{equation}
R_0=(\frac{A}{2s})^{\frac{1}{3} }N^{\frac{ 4}{9 }}
\end{equation}
and
\begin{equation}
{\cal E}_0=\frac{3A}{ 2 R_0}N^{\frac{ 4}{3 }}
\end{equation}
Because inside the ball all fermions are massless in the first
approximation $ v_{ *} \sim 0$ whereas outside  they get
large masses, they have the natural tendency to fill the ball.
They will give the stabilizing (repulsive) term in the expression
for the
total energy of the whole system which will protect from the
gravitational
collapse.
This picture will be energetically favourable until the Fermi
level $\varepsilon_{F}$
doesn't exceed the value of the fermion masses in the broken
symmetry phase.
It means that inside the ball fermions whose masses outside the
ball
are larger then inside (t-quark  etc) dominate.
Because inside the ball energy of the supersymmetric ground state
equals zero
whereas outside the ball the cosmological constant (energy of the
ground state)
also equals zero, the potential $U(x)$ describes a soliton solution
with
the bag constant $B = 0$ and only with different from zero surface
tension.
We can notice a similarity to a quark star. In this case there is
a deconfinement
phase
inside the soliton with $B \neq 0$ which is decisive for
macroscopic properties
of the star. In our case $B = 0$ and the values of the bag mass and
radius
are determined only by the surface tension.
We shall have the critical  radius $R_c$ when $R_g=R_0$. As
$R_g=\frac{2
{\cal E}}{M^2_{Pl}}$, we have  the critical fermion number $N_c$
defined as
\begin{equation}
N_c=\frac{ 1}{A}( \frac{M^2_{Pl}}{3 \sqrt{4 s^2}})^3
\end{equation}
        \begin{eqnarray*}
R_c=\sqrt{\frac{3A}{M_{Pl}}}, \ \ \ \ \ \
{\cal M}_c =\frac{1}{2} M^2_{Pl} R_c
        \end{eqnarray*}
The previous numerical parameters give
$N_c=5.27 \times 10^{69}$, $R_c=2.37 \times 10^{14}$ cm and $ M_c
=8.06 \times 10^9\ M $.
The ball mass will  depend as $N^{\frac{8}{9}}$, while the energy
of the
corresponding $N$ free particles will depend
linearly on $N$. This suggests that the ball is stable considering
the
decay of free
particles.
The proposed gauge field condensation was the modest one.
More sophisticated  condensation including $ W^{}_{\mu}$ bosons
may be also considered \cite{rm:1994}. What is very interesting and a bit
anxious is that it
breaks the electric charge conservation. Fortunately, this phase is
energetically unstable.
The current idea of a quasar is that its energy comes from the  matter
accretion on the
supermassive black holes. Nevertheless this model can not solve many
astrophysical problems associated for example with the early formation of such
massive black holes \cite{wk:1990}.
An alternative explanation of a quasar is connected with the
phenomena of  phase transitions in the early universe.
The grand unification theory predicts the sequence of  phase transitions
during the evolution of the early universe. If they are discontinued then
the bubbles of the new low temperature phase will appear during the
universe expansion. After the phase transition point the low temperature phase
will dominate and the areas of the old high temperature phase also will
form bubbles. In the presence of fermions inside, the soliton is stabilized by
surface tension term ($ \sim R^2 $).
As the result the equilibrium
configuration appears with definite mass and radius.  The comparatively
late phase transition takes place in the standard model during the
spontaneous  symmetry breaking from $ SU_L (2) \times U_Y (1) $ to $U_Q (1)$.
If such a phase transition is discontinued then the bubbles of the high
temperature phase filled  for example  with neutrinos may be produced.
In this paper it was shown that inside the bubble we have only the Dirac
neutrino
with mass of the order of the electron or quark mass. This implies that the
total lepton number is conserved inside the ball.
In the broken phase two Majorana mass eigenstates were obtained.
If we put such bubbles into the interstellar medium they may produce the
identical accretion as we expect from the supermassive black holes.
 According to Holdom \cite{bh:1987},\cite{hold:1993},\cite{hold2:1993}
a lifetime
to the value of Fermi level in the ball.
For a ball with a lifetime comparable to the age of the
Universe gravitational
interactions would
prevail and the conversion of mass could be compared with that in
the black
hole model.\\
The existence of balls can be connected with a cosmological phase
transition in the Standard Model
extended to the case with the lepton symmetry breaking.
Balls are created in the early Universe as a consequence of a
quantum tunnelling
effect or by thermal fluctuations.
Empty balls tend to shrink and disappear. In order to stabilize them
in the highsymmetric
phase Dirac neutrinos are present.
However, after the phase transition two neutrinos appear as the
 result of the see-saw
mechanism.
One of them possesses a very big mass whereas
the mass of the second neutrino is small.
It is  natural for heavy neutrinos to fall to the interior of
the ball where they are
almost massless. This is the same mechanism that leads to the
quark confinement
in the Friedberg-Lee nontopological model of hadrons. Fermions
falling into
the ball  stabilize it and cause the increase of the ball
radius
and mass.\\
The number of bubbles which  survive and their sizes depend mainly
on
the fermion density in the early Universe. Perhaps it would be
relevant to look
for any correlations with the dark matter. These balls could be good
candidates for
compact dark objects.

\section{Conclusion}
In this paper we have shown the possibility of
boson  condensation in the extreme conditions in the standard
electroweak theory.
 The first order phase transition
induced by the radiative corrections (the Coleman-Weiberg
potential) in the presence of matter was considered.
Due to t-quark mass ($\sim 174\ GeV$) a relatively high Higgs mass
($\sim 313 \ GeV$) was obtained. Only a fraction of this mass is
connected to the Coleman-Weinberg potential ($m_{CW} \sim 15\
GeV$). The model produces the first order phase transition for low
temperature ($T_c \sim 10\ GeV$).
Such a phase transition may have an astrophysical meaning, and may be
connected to the baryogenesis for the electroweak scale \cite{fr:1984}.
The realistic ball with $ M \sim 10^5
- 10^9 M_{\odot}$
and the radius $ R \sim 10^{12} - 10^{14} cm$ is obtained.

\vspace{2cm}\noindent
{\bf Acnowledgments}\\
\\
This work was written while in residence at Wisconsin of University-Madison
Department of Physics.
It is a pleasure to thank them for the warm and efficient hospitality.\\
This work was presented at the International
Symposium {\it Physics Doesn't Stop},
Madison, U.S.A., 11--13th April, 1994.\\
This research was supported in part by the U.S.~Department of Energy under
Contract No.~DE-AC02-76ER00881,  in part by the University of Wisconsin
Research Committee with funds granted by the Wisconsin Alumni Research
Foundation, and
in part by a U.S.A-Poland Maria Sk{\l }odowska-Curie Joint
Fund II.\\

\newpage
\setlength{\unitlength}{0.240900pt}
\ifx\plotpoint\undefined\newsavebox{\plotpoint}\fi
\sbox{\plotpoint}{\rule[-0.500pt]{1.000pt}{1.000pt}}%


\noindent
Figure 1.
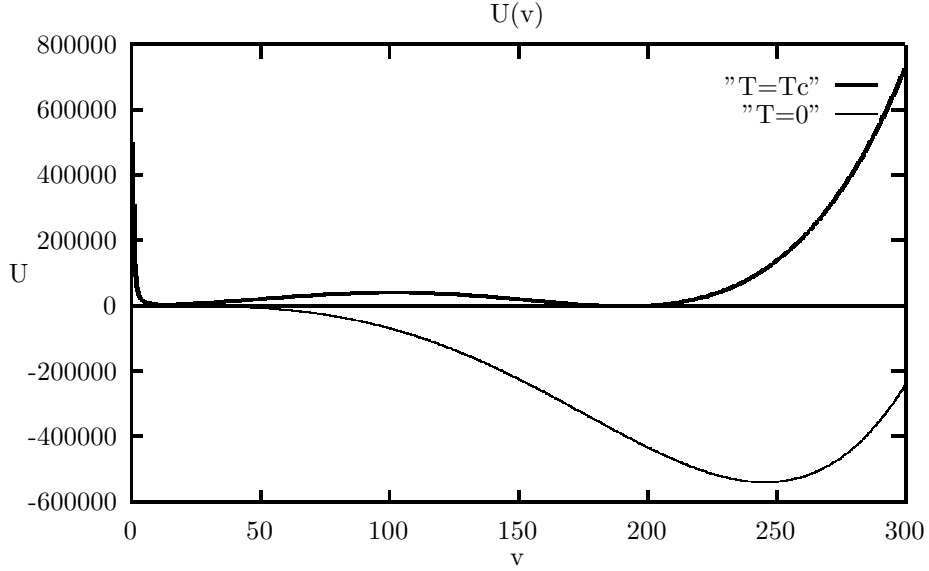
The potential $ U(x) $ for external fermions  densities $\rho_Z\ \ne \ 0$
for temperature $T=0$ and $T=T_c$.\\

\vspace{1in}

\begin{table}[h]
\begin{center}
\caption{The value of cosmological constant B, phase transition
temperature $T_c$ and Higgs total mass $m_H$ for different
parameters of the Coleman-Weinberg Higss mass $m_{cw}$}
\label{tabel}
\vspace{0.5cm}
\begin{tabular}{|c|c|c|c|}
\hline
\hline
$B $ & $T_c $ & $m_H$ & $m_{CW}$  \\ \hline
\hline
$-9.2\ 10^5 \ (GeV)^4 $ & $10.0 \ GeV$ & $ 313.0\ GeV$ &$15 \ GeV$   \\ \hline
$-2.5\ 10^6 \ (GeV)^4 $ & $15.9 \ GeV$ & $ 327.7\ GeV$ &$25.9
\ GeV$   \\ \hline
$-5.6\ 10^6 \ (GeV)^4 $ & $22.9 \ GeV$ & $ 352.7\ GeV$ &$37.2 \ GeV$   \\
\hline
\hline
\end{tabular}
\end{center}
\end{table}
\end{document}